© *A. S. Yurkov*
Omskiy Nauchno Issledovatelskiy Institut Priborostroeniya, Institute of Radiophysics and Physical Electronics OSC SB RAS, Omsk, Russian Federation

## TO THE THEORETICAL DESCRIPTION OF THE LOSSES IN INDUCTANCE COILS

The article discusses a theory that allows to describe losses in conductors taking into account the skin effect and the proximity effect. The developed theory is applied to inductance coils. In contrast to the classical Butterworth theory the condition of the strong skin effect, which is frequent in practice, is introduced from the very beginning. This approach makes possible to obtain extremely simple formulas, which show that Q-factor of a coil is determined mainly by the diameter of its winding.

© *А. С. Юрков*
Омский научно-исследовательский институт приборостроения, Институт радиофизики и физической электроники ОНЦ СО РАН, Омск, Российская Федерация

## К ТЕОРЕТИЧЕСКОМУ ОПИСАНИЮ ПОТЕРЬ В КАТУШКАХ ИНДУКТИВНОСТИ

Рассматривается теория, позволяющая описывать потери в проводниках с учетом скин-эффекта и эффекта близости. Развитая теория применяется к катушкам индуктивности. В отличие от классической теории Баттерворта, с самого начала вводится часто встречающееся на практике условие сильного скин-эффекта. Этот подход позволяет получить крайне простые формулы, которые показывают, что добротность катушки определяется в основном диаметром ее намотки.

### 1. Introduction

Inductance coils (hereafter just coils for brevity) are important elements of many radio devices. Therefore, the need to calculate their parameters is beyond the doubt. Omitting the parameters of the second plan, such as parasitic capacitance, one can say that the main electrical parameters are inductance $L$ and Q-factor $Q$. The calculation of the Q-factor is equivalent in essence to the calculation of the active resistance $r$, related to $Q$ and $L$ by the simple and well-known equation.

The calculation of the inductance of the coil, at least within the accuracy usual for engineering practice, does not cause any serious problem. Calculation of the losses, characterized by Q-factor or active resistance, is a more complicated problem. In principle, presently this problem can be solved by direct computer simulation of the electromagnetic field using one or another software package. However, it is quite labor-consuming. Attention should also be paid to large computing resources needed for such modeling. This is why the original specialized numerical methods, that make such modeling more effective, are still offered in the scientific literature (see, for example, [1]).

Despite the fact that computer simulation provides a direct solution to the problem, the presence of simple analytical formulas, even if they are not of high accuracy, is highly desirable. The fact is that the dependences of the calculated value on the design parameters of the coil are clearly visible from such formulas, this is very useful for appropriate choice of these parameters. In addition, the calculation by such formulas is matter of a few minutes. So that, an approach based on analytical calculations and aimed at getting the simple formulas is of particular interest.

The theory of losses in coils began with the papers by Butterworth published at the beginning of the 20th century [2–4]. Butterworth's approach is based on a fairly straightforward solution to the electrodynamic problem about a cylindrical wire taking into account so-called proximity effect. It was done by decomposing the electromagnetic field in a series of Bessel functions. Although such approach is accurate at first glance, further rather rough approximations were made anyway. In fact, this approach is meaningful mainly in a rather special case, when the skin layer thickness is of the same order of magnitude or larger than the diameter of the wire. Nevertheless, the approach based on the expansion in Bessel functions still dominates in the literature (see, for example, [5]).

It is shown in this paper that in the frequent case, when the skin layer thickness is much smaller than the wire diameter, the corresponding electrodynamic problem can be solved by much simpler means than the direct decomposition in Bessel functions. The result is extremely simple formulas (much simpler than the Butterworth's formulas), which are very convenient for approximate estimates in engineering practice. It should be noted that within the framework of the proposed approach the accuracy of the calculation can be improved if the approximation for the external field, adopted in this paper, is improved in one way or another (this will become clear from what follows). It is important that such an improvement does not affect, however, the basis of the proposed approach. Therefore, matter considered in this paper has also a theoretical meaning. At the same time, improving the accuracy quite complicates final formulas. Therefore, focusing on practice as well in connection with a limited volume of the paper, only the main approximation which gives the simplest result is considered here.

## 2. Physical nature of the losses in inductance coils

We restrict ourselves to the case of coils without a magnetic core. In this case four mechanisms of loss having different physical nature can be distinguished for the inductance coils:

1) radiation losses;
2) dielectric loss in the frame material etc.;

3) environment losses due to the presence of surrounding objects in which currents are induced;

4) losses due to ohmic resistance of the wire.

Each of these mechanisms can be characterized by a partial Q-factor $Q_i$ (index $i$ numbers the above mechanisms) which the coil would have if there were no other mechanisms. The total Q-factor of the coil $Q$ can be determined by the obvious formula:

$$\frac{1}{Q} = \sum_i \frac{1}{Q_i}. \qquad (1)$$

Radiation losses in order of magnitude can be estimated using the well-known Chu equation (see, for example, [6]):

$$Q_{rad} = \frac{1}{(ka)^3}, \qquad (2)$$

where $k$ is the wave number equal to $2\pi$ divided by the length of the corresponding wave, $a$ is the characteristic size of the system (radius of the covering sphere). Substituting into this equation the values usual for practice it is easy to verify that the corresponding value of $Q_{rad}$ is equal to several thousands at least (but only at fairly high frequencies). Since total Q-factor obtained in practice is usually much less, it can be concluded that this loss mechanism is not predominant and can be neglected.

Q-factor associated with dielectric losses in any case is greater than the reverse tangent of dielectric losses. With high-quality dielectrics and even more so with frameless winding it turns out that this mechanism is usually not the main one.

The environment losses obviously depends on the configuration of the environment. Since this configuration can be very different it is not possible to make general estimates. Often the environment losses can be neglected, this is clearly seen if one measure Q-factor of the same coil in different environments. This is why we do not consider this type of losses.

Thus, the main loss mechanism which determines total $Q$ corresponds to ohmic losses in the wire, from which the coil is made. From a practical point of view, it can be assumed that there is this mechanism only (excluding, of course, unusual cases, for example, the case of a frame made from low-quality dielectric).

One might think that the magnitude of the ohmic loss resistance and the corresponding Q-factor can be calculated as for a conventional wire using the well-known formula that takes into account, of course, the skin effect. However, practice shows that such calculation yields greatly overestimated, in comparison with reality, value of $Q$. The reason for this is the fact that when the wire is surrounded by other wires with the same current (just the case of a coil), the current is distributed non-homogeneously around the circumference of the wire. So that, the effective surface that carries the current becomes smaller and the resistance increases. This is so-called proximity effect. In fact, the further consideration is aimed at the finding simple analytical dependences describing this effect including it in the inductance coils.

### 3. Current distribution over the surface of cylindrical wire and energy absorption in the conditions of strong skin effect

Almost always the diameter of the wire, which is used in the inductance coil, is much smaller than the diameter of the winding of this coil. So that, considering a small part of the wire it is quite possible to consider it as a straight-line cylinder. In turn, the thickness of the skin layer, in which current flows, is usually much smaller than the diameter of the wire (at least at frequencies of SW band and above). Thus, we can assume that the current is purely surface one and that it flows along the cylindrical surface in the direction of the axis of the cylinder. From what follows, it is clear that in order to find the current distribution along the wire circumference, first of all, it is necessary to calculate the magnetic field strength created by this surface current (hereafter this and other magnetic field strengths are called as magnetic field or even as field for short).

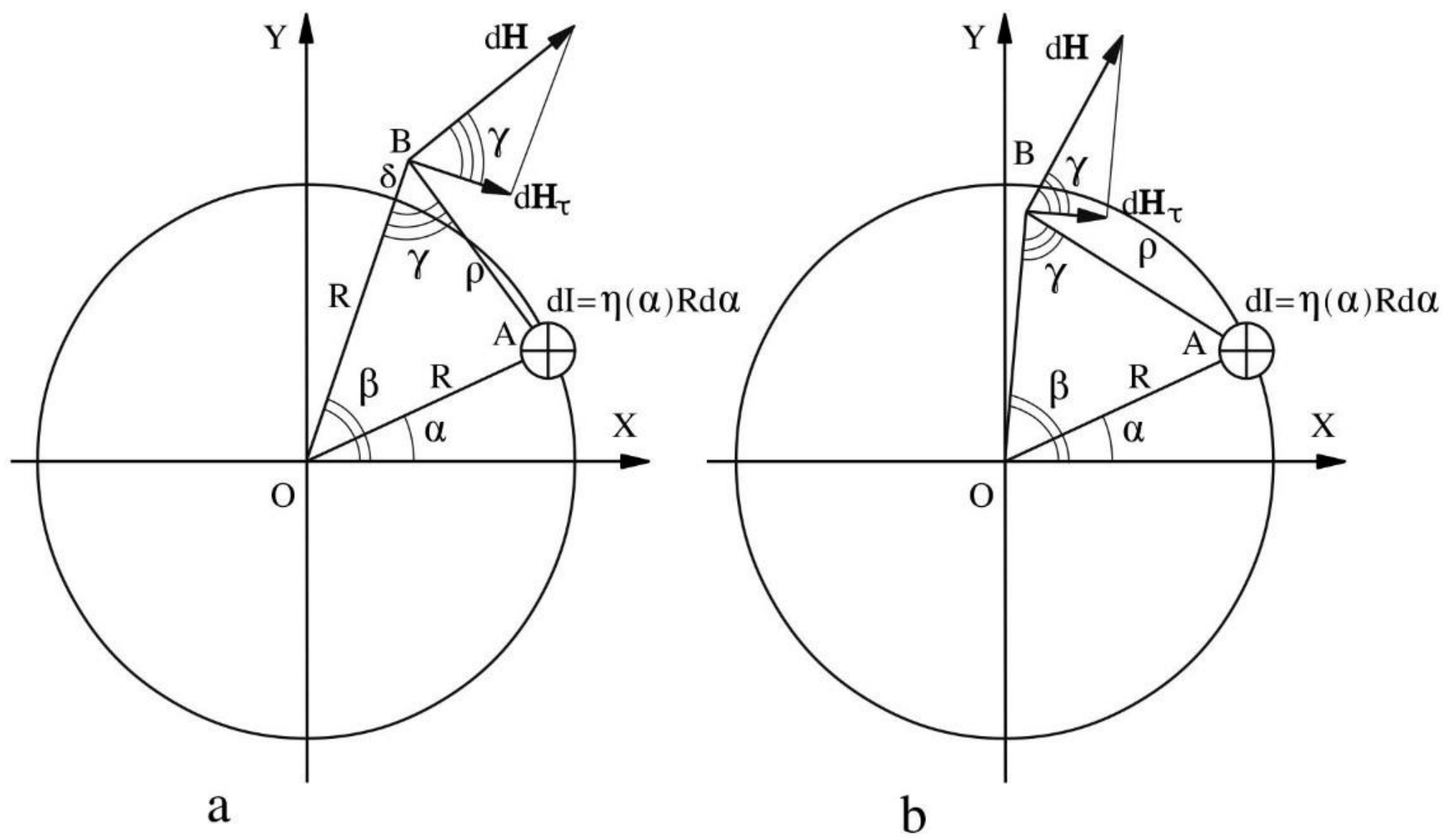

**Fig. 1**. The geometry that determines the field strength produced by elementary current

We use a cylindrical coordinate system with the axis $Z$ coinciding with the axis of the cylinder. Let $R$ be the radius of the cylinder, $\eta(\alpha)$ be the surface current density (directed parallel to the cylinder axis) depending on the polar angle $\alpha$ (there is no dependence on the $Z$ coordinate, as assumed here and further). The magnetic field $H(\beta)$ at a point characterized by the polar angle $\beta$ is defined by the following obvious equation:

$$H(\beta) = \frac{1}{2\pi}\int_0^{2\pi} \frac{\eta(\alpha) R d\alpha}{\rho}, \tag{3}$$

where $\rho$ is the distance from the current element $\eta(\alpha)Rd\alpha$ to the point at which the field is calculated.

We are interested in the field on the surface of the cylinder. However, the radial coordinate of the point, at which the field is calculated, cannot simply be set equal to $R$ because a singularity appears in the integral. So that, analysis using a limiting transition is required. To perform it, we calculate the field at small distance $\delta$ from the surface of the cylinder and only after a selection of the singularity we tend $\delta$ to zero.

It is necessary to distinguish between the case of calculating the field outside the cylinder (see Fig.1a) and the case of calculating the field inside the cylinder (see Fig. 1, b). Due to the fact that the current flows on the surface of the cylinder, the magnetic field undergoes a jump and it is different on the different sides of this surface. However, much of the calculations can be done without separate calculation of the field outside and inside the cylinder. We can assume that when calculating the field outside the cylinder the value of $\delta$ is positive and when calculating the field inside the cylinder it is negative (this is further implied). With such an agreement, using the cosine theorem for the $OAB$ triangle (see Fig.1) in both cases we can write the same equation:

$$R^2 = \rho^2 + (R+\delta)^2 - 2\rho(R+\delta)\cos\gamma. \quad (4)$$

Further elementary transformations yield following:

$$\rho\cos\gamma = \frac{\rho^2}{2(R+\delta)} + \frac{2R+\delta}{2(R+\delta)}\delta. \quad (5)$$

The usefulness of equation (5) is that, as is seen below, we need not total magnetic field $H$ but only its component $H_\tau$ which is tangential to the surface of the cylinder. The tangential component is related to total field $H$ through the cosine of the same angle $\gamma$ (angles with mutually perpendicular sides, see Fig.1). It reads $H_\tau = H\cos\gamma$. Thus, using (3), multiplying both the numerator and the denominator in the integral by the same factor $\rho$, going from $H$ to $H_\tau$ and applying (5) we get the following:

$$H_\tau(\beta) = \frac{1}{4\pi(R+\delta)}\int_0^{2\pi}\eta(\alpha)Rd\alpha + \frac{R(2R+\delta)}{4\pi(R+\delta)}\int_0^{2\pi}\frac{\delta}{\rho^2}\eta(\alpha)d\alpha. \quad (6)$$

The first integral in this equation reduces to the total current $I$ flowing over the surface of the cylinder. The limit $\delta \to 0$ in this term does not require special analysis. As a result, the first term is equal $I/(4\pi R)$.

The singularity mentioned above is present in the second term in the right-hand side of (6). The treatment of this singularity is based on the fact that in the limit $\delta \to 0$ term $\delta/\rho^2$, as a function of $\alpha$, turns out to be $\delta$-function up to a factor. For small but finite values of $\delta$ this function has a very narrow, in the limit infinitely narrow, peak near the point $\alpha = \beta$ (when the distance $\rho$ is small). Therefore, the second integral in the right-hand side of (6) is determined only by this small neighborhood within which the slowly varying function $\eta(\alpha)$ can be considered as a constant and can be removed from the integral. As a result, taking into account what was said above about the first term we get the following:

$$H_\tau(\beta) = \frac{I}{4\pi R} + \eta(\beta) \lim_{\delta \to 0} \left[ \frac{R(2R+\delta)}{4\pi(R+\delta)} \int_0^{2\pi} \frac{\delta}{\rho^2} d\alpha \right]. \qquad (7)$$

To calculate the integral under the limit in this equation, first one need to find an explicit expression for $\rho^2$. Using the cosine theorem again for the same triangle, but taking another angle, we write the following:

$$\rho^2 = R^2 + (R+\delta)^2 - 2R(R+\delta)\cos(\beta - \alpha). \qquad (8)$$

Substituting this expression into the integral of interest and making a standard replacement of the integration variable by tangent of the half angle, this integral can be calculated easy. As a result, turning to the limit not constituting serious problem now, we get the following very simple expression:

$$H_\tau(\beta) = \frac{I}{4\pi R} \pm \frac{1}{2}\eta(\beta), \qquad (9)$$

where the plus sign corresponds to the field on the outer surface of the cylinder while the minus sign should be used on the inside surface. Note that in the particular case of the uniformly distributed current this expression for external surface yields a standard equation obtained by circulation theorem, while it yields zero for the internal one (as it should be by the same theorem).

Using (9) one can find the current distribution over the surface of the wire under the conditions when this wire is surrounded by other wires carrying the same current. To do this just note that inside the wire (under the skin layer that is, in our approximation, on the inside surface of the cylinder) the sum of the field obtained by (9) and the tangential component of the external field $H_\tau^{ext}$ created by other wires obviously should vanish. Thus, we get the following:

$$\eta(\beta) = \frac{I}{2\pi R} + 2H_\tau^{ext}(\beta). \qquad (10)$$

Equation for the tangential component of the total field on the outer surface obtained by excluding $\eta(\beta)$ is useful also. It reads as follows:

$$H_\tau^{tot}(\beta) = \frac{I}{2\pi R} + 2H_\tau^{ext}(\beta). \qquad (11)$$

Obtained equation (11) allows us to write a general expression for the power $P$ absorbed by cylindrical surface of the part of wire having length $l_W$. One should use the well-known Leontovich boundary conditions (which are exact in the limit of thin skin layer) to express longitudinal electric field $E_Z$ in terms of $H_\tau^{tot}$. Then one should form the real part of Poynting vector normal to the surface and integrate it over considered part of the surface. The result is the following:

$$P=\frac{1}{2}(\mathrm{Re}W)l_W R\int_0^{2\pi}\left|H_\tau^{tot}(\beta)\right|^2 d\beta=\frac{1}{2}(\mathrm{Re}W)l_W R\int_0^{2\pi}\left|\frac{I}{2\pi R}+2H_\tau^{ext}(\beta)\right|^2 d\beta. \quad (12)$$

Here $W$ is surface impedance which is defined as follows (see, for example, [7]):

$$W=(1+i)\sqrt{\frac{\omega\mu_0}{2\sigma}}=\frac{1+i}{\sigma\Delta}, \quad (13)$$

where $\omega$ is a circular frequency, $\mu_0$ is the magnetic constant (it is assumed that the wire material is nonmagnetic), $\sigma$ is a conductivity of the wire material, $\Delta=\sqrt{2/(\omega\mu_0\sigma)}$ is the skin layer thickness.

When calculating losses in a coil, field $H_\tau^{ext}(\beta)$ is created by parts of the same wire as the considered part. All of these parts of the wire carry the same current $I$. Therefore, $H_\tau^{ext}(\beta)$ is proportional to $I$. Let us write this down by entering a real function $\psi(\beta)$, which is defined below, as follows:

$$H_\tau^{ext}(\beta)=I\psi(\beta). \quad (14)$$

Using this as well as the above expression for $W$ equation (12) is converted to following:

$$P=\frac{1}{2}|I|^2\frac{l_W R}{\sigma\Delta}\int_0^{2\pi}\left[\frac{1}{2\pi R}+2\psi(\beta)\right]^2 d\beta. \quad (15)$$

This equation is easy recognized as usual expression for active power $P=(1/2)|I|^2 r$, where $r$ is active resistance. Hence we get the general expression for resistance per unit length:

$$\frac{r}{l_W}=\frac{R}{\sigma\Delta}\int_0^{2\pi}\left[\frac{1}{2\pi R}+2\psi(\beta)\right]^2 d\beta. \quad (16)$$

This completes the construction of the general theory of ohmic losses in a cylindrical wire under the conditions of the proximity effect and the strong skin effect. Attention should be paid to the fact that the general theory is built using only the approximation of the strong skin effect. There is no need to use any expansion in series and additional approximations. This is a significant advantage over the classical Butterworth theory. Of course this advantage is due to the fact that we have restricted ourselves to the case of the strong skin effect. The limitation is not too important because the situation of the strong skin effect is typical in radio engineering (as opposed to electrical engineering). Only at frequencies corresponding to medium waves and below this limitation can be significant and even then only in the case of a wire of small diameter.

Further one can apply the developed theory to specific situations. This involves the calculation or some approximation of the function $\psi(\beta)$. The simplest approximation for the case of a single-layer coil and the results obtained are discussed in the next section.

## 4. Application of the general theory for describing losses in a single-layer inductance coil

To apply the general theory developed above to the specific case of inductance coils one need to find one or another approximation for function $\psi(\beta)$. It means an approximation of the magnetic field external to some part of the wire. In the simplest case this field can be approximately considered as homogeneous. Note that this approximation is also used in the paper by Butterworth [2]. It is obvious that the field with high accuracy is homogeneous if the radius of the wire is much smaller than the distance between the turns of the coil. However, practice shows that this approximation works satisfactorily in the case of larger radius of the wire (see the discussion of this issue in [3]).

Further, with the aim of obtaining a simple final formula, we are forced to restrict ourselves to the case of a single-layer coil. First, we consider the case of a sufficiently long coil (the corresponding criterion determining which length is sufficient is obtained below). For this case the field $H_\tau^{ext}$ can be estimated basing on the following reasoning. It is well known that in a long coil the field inside it is equal to $I/s$, where $s$ is the winding step. Outside the coil the field is zero. Therefore, using linear interpolation we conclude that the field on the coil turns is approximately equal to the arithmetic mean that is $I/(2s)$. Further, remembering that near small portion of the wire the field is approximately considered as homogeneous, for tangential component of this field the dependence on angle is reduced to cosine. The result is the following:

$$\psi(\beta) = \frac{\cos\beta}{2s}. \tag{17}$$

Substituting this expression in (16) and completing elementary integration we obtain the following equation for active resistance per unit length of the wire:

$$\frac{r}{l_W} = \frac{1+(2\pi^2 R^2)/s^2}{2\pi\sigma R\Delta}. \tag{18}$$

Equation (18) essentially solves the problem of describing the losses in the long single-layer coil. But the use of it is extremely inconvenient. Therefore, we turn to Q-factor. Inductance is expressed by the standard formula for a long coil:

$$L = \frac{\mu_0 \pi D^2 N^2}{4l}. \tag{19}$$

Here $N$ is the number of turns, $D$ is the winding diameter, $l$ is the length of the winding. From the radius of the wire $R$ we turn to its diameter $d$ and take into account that $\omega\mu_0\sigma = 2/\Delta^2$. Then most of the parameters cancels and we obtain following equation:

$$Q = \frac{D}{\Delta\sqrt{2}} \cdot \frac{x}{1+x^2}, \qquad (20)$$

where

$$x = \frac{\pi d}{s\sqrt{2}}. \qquad (21)$$

Usually it is necessary to make Q-factor as large as possible. Therefore, it makes sense to optimize (20) with respect to $x$. By this way following equation is obtained for a coil with an optimal ratio of the winding step and the wire diameter (it means that $x=1$):

$$Q_{opt} = \frac{D}{\Delta} \cdot \frac{1}{2\sqrt{2}} \approx 0.35 \cdot \frac{D}{\Delta}. \qquad (22)$$

Optimum is obtained when $d = (s\sqrt{2})/\pi \approx 0.45s$.

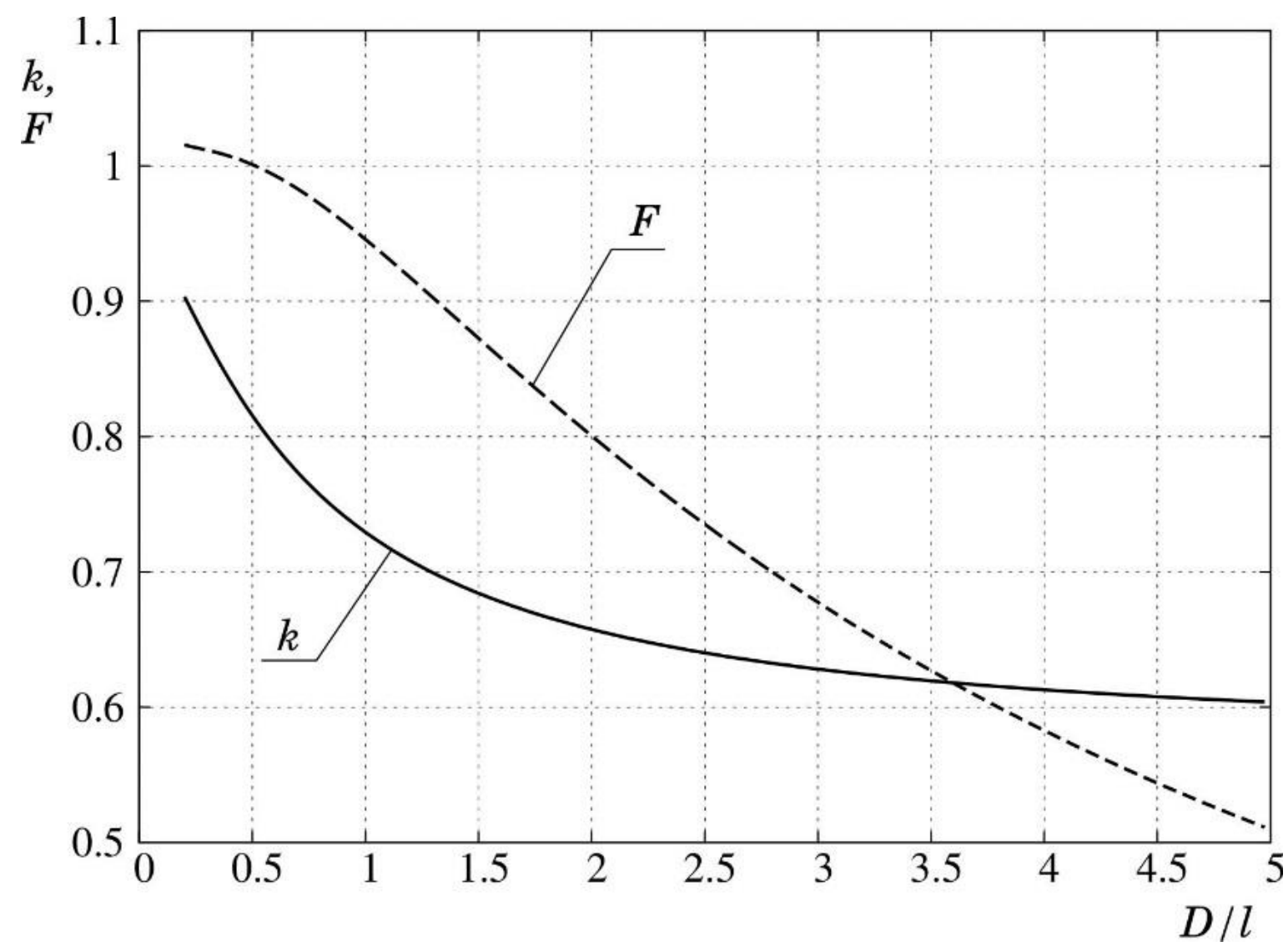

**Fig. 2**. The results of the calculations by numerical methods. Coefficient $k$ (solid line) and factor $F$ (dashed line) depending on the ratio of winding diameter and winding length $D/l$

It is more difficult to determine Q-factor for a coil with small length. To do this we add the correction coefficient $k$ into the ratio used above between the field, the winding step and the current:

$$H_{\tau}^{ext} = k \cdot \frac{I}{2s}. \qquad (23)$$

Getting a simple expression for $k$ is not possible. But it can be shown that in a reasonable approximation this coefficient depends only on the ratio between the diameter and the length of the coil. In Fig. 2 the solid line shows the result of the calculation of this coefficient by numerical methods.

For coils of small length the inductance should be determined accordingly. It is well known that for the coils used in practice Wheeler formula [8] gives good accuracy. This formula, which is slightly modified here, is as follows:

$$L=\frac{\mu_0 \pi D^2 N^2}{4l}\cdot\frac{1}{1+0.45(D/l)}. \quad (24)$$

Further, repeating the same calculations as described above, but now taking into account the coefficient $k$ and the additional factor $1/(1+0.45D/l)$ that is present in the Wheeler formula, we get the following:

$$Q=\frac{D}{\Delta\sqrt{2}}\cdot F\cdot\frac{x}{1+x^2}, \quad (25)$$

where

$$F=\frac{1}{k(1+0.45(D/l))}, \quad (26)$$

$$x=\frac{\pi dk}{s\sqrt{2}}. \quad (27)$$

The graph of the dependence of $F$ on $D/l$ is shown in Fig. 2 by the dashed line.

From the results of numerical calculations presented in Fig. 2 we can draw the following conclusions. First, the more rigorous numerical calculation confirms that for long coils $H^{ext}\approx I/(2s)$, as it is taken above. Indeed, for small $D/l$, which corresponds to the long coils, the correction coefficient $k$ tends to unity. Secondly, at the optimum ratio of the wire diameter and the winding step (it means that $x=1$), Q-factor of the short coil is less than that of the long coil (the additional factor $F$ is significantly less than unity). Third, the optimal wire diameter for short coils is somewhat larger than for long ones (the optimality condition $x=1$ implies that $d=(s\sqrt{2})/(k\pi)$). At the same time, if the coil length is equal to or more than the winding diameter (which is reasonable to do aiming to achieve maximum $Q$), then $F\approx 1$ and it is quite possible to use simpler formulas derived above for the case of long coils.

## 5. Conclusion

In this paper we develop the theoretical description of losses in conductors which takes into account the proximity effect and which is radically simpler than in the classical papers by Butterworth. The achieved simplicity of the theory is due to the fact that from the very beginning the condition of the strong skin effect is accepted (the skin layer thickness is much less than the wire diameter). Of course this condition leads to the fact that the theory developed here has a more limited scope than the Butterworth theory. However, the situation of the strong skin effect occurs very often, in radio engineering even more often than the reverse situation. So that, the theory considered here has a very wide area of applicability.

The application of the considered theory to the calculation of Q-factor of single-layer inductance coils without a core led to extremely simple equations. Especially simple equations are obtained for the case of long coils. As further research has shown, from the point of view of calculating the quality factor the long coils are those whose winding length is equal to or greater than the winding diameter. It is significant that short coils, as it turned out, have lower $Q$. So that, their use in most cases is not desirable. However, if it becomes necessary to use short coils then to calculate their Q-factor one only need to enter additional correction factors into equations. For this factors the corresponding graphs are given above.

An interesting result of this work is also the obtained condition of optimality of the wire diameter that the coil is wound. This diameter should be 45 percent (roughly speaking half) of the winding step. It turns out that if the diameter of the wire is optimal then the quality factor of the coil is determined only by the diameter of its winding (it is assumed that the thickness of the skin layer is fixed, the material of the wire is usually fixed, the frequency here is also considered fixed). Strictly speaking, this statement is true only for long coils, for short ones there is also a dependence on the ratio of the diameter and length of the winding, but this additional dependence is rather weak.

## Information about the author

***Yurkov Alexander Sergeevich*** – PhD, JSC "ONIIP", Senior researcher Institute of Radiophysics and Physical Electronics OSC SB RAS. E-mail: trs@oniip.ru.